\documentclass[%
 reprint,
superscriptaddress,
 amsmath,amssymb,
 aps,
]{revtex4-2}

\usepackage{graphicx}
\usepackage{dcolumn}
\usepackage{bm}
\usepackage{upgreek}

\begin{document}

\preprint{APS/123-QED}

\title{Improved limit on the effective electron neutrino mass with the ECHo-1k experiment}
\author{F.~Adam}\affiliation{Institute for Astroparticle Physics, Karlsruhe Institute of Technology, Germany}\affiliation{Institute for Micro- and Nanoelectronic Systems, Karlsruhe Institute of Technology, Germany}
\author{F.~Ahrens}\altaffiliation[Presently at ]{Fondazione Bruno Kessler, (FBK), I-38123 Trento, Italy and INFN- TIFPA, I-38123 Trento, Italy}\affiliation{Kirchhoff Institute for Physics, Heidelberg University, Germany}
\author{L.~E.~Ardila~Perez}\affiliation{Institute for Data Processing and Electronics, Karlsruhe Institute of Technology, Germany}
\author{M.~Balzer}\affiliation{Institute for Data Processing and Electronics, Karlsruhe Institute of Technology, Germany}
\author{A.~Barth}\affiliation{Kirchhoff Institute for Physics, Heidelberg University, Germany}
\author{D.~Behrend-Uriarte}\affiliation{Kirchhoff Institute for Physics, Heidelberg University, Germany}
\author{S.~Berndt}\affiliation{Department of Chemistry - TRIGA Site, Johannes Gutenberg University Mainz, Germany}\affiliation{Institute of Physics, Johannes Gutenberg University Mainz, Germany}
\author{K.~Blaum}\affiliation{Max Planck Institute for Nuclear Physics, Heidelberg, Germany}
\author{F.~W.~H.~B\"ohm}\affiliation{Kirchhoff Institute for Physics, Heidelberg University, Germany}
\author{M.~Bra{\ss}}\affiliation{Institute for Theoretical Physics, Heidelberg University, Germany}
\author{L.~Calza}\affiliation{Kirchhoff Institute for Physics, Heidelberg University, Germany}
\author{K.~Chrysalidis}\affiliation{ISOLDE, CERN, Geneve, Switzerland/France}
\author{M.~Door}\altaffiliation[Presently at ]{Laboratoire Kastler Brossel, Sorbonne University, CNRS, ENS-PSL University, Collège de France, Paris, France.}\affiliation{Max Planck Institute for Nuclear Physics, Heidelberg, Germany}
\author{H.~Dorrer}\affiliation{Department of Chemistry - TRIGA Site, Johannes Gutenberg University Mainz, Germany}
\author{Ch.~E.~D\"ullmann}\affiliation{Department of Chemistry - TRIGA Site, Johannes Gutenberg University Mainz, Germany}\affiliation{Helmholtz Institute Mainz, Mainz, Germany}\affiliation{GSI Helmholtz Centre for Heavy Ion Research GmbH, Darmstadt, Germany}
\author{K.~Eberhardt}\affiliation{Department of Chemistry - TRIGA Site, Johannes Gutenberg University Mainz, Germany}\affiliation{Helmholtz Institute Mainz, Mainz, Germany}
\author{S.~Eliseev}\affiliation{Max Planck Institute for Nuclear Physics, Heidelberg, Germany}
\author{C.~Enss}\affiliation{Kirchhoff Institute for Physics, Heidelberg University, Germany}\affiliation{Institute for Data Processing and Electronics, Karlsruhe Institute of Technology, Germany}
\author{P.~Filianin}\affiliation{Max Planck Institute for Nuclear Physics, Heidelberg, Germany}
\author{A.~Fleischmann}\affiliation{Kirchhoff Institute for Physics, Heidelberg University, Germany}
\author{R.~Gartmann}\affiliation{Institute for Data Processing and Electronics, Karlsruhe Institute of Technology, Germany}
\author{L.~Gastaldo}\email[Corresponding author: ]{Gastaldo@kip.uni-heidelberg.de}\affiliation{Kirchhoff Institute for Physics, Heidelberg University, Germany}\
\author{M.~Griedel}\altaffiliation[Presently at ]{Institute for Quantum Material and Technologies, Karlsruhe Institute of Technology, Germany}\affiliation{Kirchhoff Institute for Physics, Heidelberg University, Germany}
\author{A.~G\"oggelmann}\affiliation{Physics Institute, University of T\"ubingen, Germany}
\author{R.~Hammann}\altaffiliation[Presently at ]{Max Planck Institute for Nuclear Physics, Heidelberg, Germany}\affiliation{Kirchhoff Institute for Physics, Heidelberg University, Germany}
\author{R.~Hasse}\affiliation{Institute of Physics, Johannes Gutenberg University Mainz, Germany}
\author{M.W.~Haverkort}\affiliation{Institute for Theoretical Physics, Heidelberg University, Germany}
\author{S.~Heinze}\affiliation{Institute for Theoretical Physics, Heidelberg University, Germany}
\author{D.~Hengstler}\affiliation{Kirchhoff Institute for Physics, Heidelberg University, Germany}
\author{R.~Jeske}\affiliation{Kirchhoff Institute for Physics, Heidelberg University, Germany}
\author{J.~Jochum}\affiliation{Physics Institute, University of T\"ubingen, Germany}
\author{K.~Johnston}\affiliation{ISOLDE, CERN, Geneve, Switzerland/France}
\author{N.~Karcher}\affiliation{Institute for Data Processing and Electronics, Karlsruhe Institute of Technology, Germany}
\author{S.~Kempf}\affiliation{Institute for Micro- and Nanoelectronic Systems, Karlsruhe Institute of Technology, Germany}\affiliation{Institute for Data Processing and Electronics, Karlsruhe Institute of Technology, Germany}
\author{T.~Kieck}\affiliation{GSI Helmholtz Centre for Heavy Ion Research GmbH, Darmstadt, Germany}\affiliation{Helmholtz Institute Mainz, Mainz, Germany}
\author{U.~K\"oster}\affiliation{Institut Laue-Langevin, Grenoble, France}
\author{N.~Kovac}\altaffiliation[Presently at ]{Institute for Astroparticle Physics, Karlsruhe Institute of Technology, Germany}\affiliation{Kirchhoff Institute for Physics, Heidelberg University, Germany}
\author{N.~Kneip}\affiliation{Institute of Physics, Johannes Gutenberg University Mainz, Germany}
\author{K.~Kromer}\affiliation{Max Planck Institute for Nuclear Physics, Heidelberg, Germany}
\author{F.~Mantegazzini}\altaffiliation[Presently at ]{Fondazione Bruno Kessler, (FBK), I-38123 Trento, Italy and INFN- TIFPA, I-38123 Trento, Italy}\affiliation{Kirchhoff Institute for Physics, Heidelberg University, Germany}
\author{{\textdagger}B.~A.~Marsh}\affiliation{ISOLDE, CERN, Geneve, Switzerland/France}
\author{M.~Merstorf}\affiliation{Institute for Theoretical Physics, Heidelberg University, Germany}
\author{T.~Muscheid}\affiliation{Institute for Data Processing and Electronics, Karlsruhe Institute of Technology, Germany}
\author{M.~Neidig}\affiliation{Institute for Micro- and Nanoelectronic Systems, Karlsruhe Institute of Technology, Germany}
\author{Y.~N.~Novikov}\affiliation{Max Planck Institute for Nuclear Physics, Heidelberg, Germany}
\author{R.~Pandey}\affiliation{Kirchhoff Institute for Physics, Heidelberg University, Germany}
\author{A.~Reifenberger}\affiliation{Kirchhoff Institute for Physics, Heidelberg University, Germany}
\author{D.~Richter}\affiliation{Kirchhoff Institute for Physics, Heidelberg University, Germany}
\author{A.~Rischka}\affiliation{Max Planck Institute for Nuclear Physics, Heidelberg, Germany}
\author{S.~Rothe}\affiliation{ISOLDE, CERN, Geneve, Switzerland/France}
\author{O.~Sander}\affiliation{Institute for Data Processing and Electronics, Karlsruhe Institute of Technology, Germany}
\author{R.~X.~Sch\"ussler}\affiliation{Max Planck Institute for Nuclear Physics, Heidelberg, Germany}
\author{S.~Scholl}\affiliation{Physics Institute, University of T\"ubingen, Germany}
\author{Ch.~Schweiger}\affiliation{Max Planck Institute for Nuclear Physics, Heidelberg, Germany}
\author{C.~Velte}\affiliation{Kirchhoff Institute for Physics, Heidelberg University, Germany}
\author{M.~Weber}\affiliation{Division V, Karlsruhe Institute of Technology, Germany}
\author{M.~Wegner}\affiliation{Institute for Data Processing and Electronics, Karlsruhe Institute of Technology, Germany}\affiliation{Institute for Micro- and Nanoelectronic Systems, Karlsruhe Institute of Technology, Germany}
\author{K.~Wendt}\affiliation{Institute of Physics, Johannes Gutenberg University Mainz, Germany}
\author{T. Wickenh\"auser}\affiliation{Kirchhoff Institute for Physics, Heidelberg University, Germany}
 

\collaboration{ECHo Collaboration}

\date{\today}

\begin{abstract}
The effective electron neutrino mass can be determined by analyzing the endpoint region of the $^{163}$Ho electron capture spectrum, provided a measurement with high energy resolution and high statistics using calorimetric techniques. Here, the Electron Capture in $^{163}$Ho collaboration, ECHo, presents an analysis of the most precise $^{163}$Ho spectrum currently available, obtained with the ECHo-1k experiment and comprising about 200 million events. A very low background rate of $B=9.1(1.3)\times 10^{-6}$ /eV/pixel/day was achieved allowing for a reliable analysis of the endpoint region. The derived endpoint energy $Q = 2862(4)$ eV is in excellent agreement with the one independently determined via Penning-trap mass spectrometry of $Q=2863.2(6)$ eV \cite{PTMS2024}. The upper limit of the effective electron neutrino mass is improved by almost a factor 2 compared to the lowest current value \cite{Holmes_25}, reaching $m_{\nu_\mathrm{e}} < 15 $ eV/c${^2}$ (90\% credible interval).

\end{abstract}

\maketitle



The determination of the neutrino mass scale will give guidance for extending the present Standard Model of Particle Physics \cite{MassModel} and enable precision tests of the $\Lambda$CDM Cosmological Model \cite{nucosmo}. Three different observables, which are functions of the neutrino mass eigenvalues $m_{\mathrm{i}}$, are investigated to fix the neutrino mass scale: the sum of the neutrino mass eigenvalues from the analysis of the distribution of matter in the Universe, the effective Majorana mass which can be derived if neutrinoless double beta decay is observed and the effective electron neutrino mass \cite{Cosmo, Jiang_2025, double, RevModPhys.95.025002, FORMAGGIO20211}. 
The derivation of the effective electron neutrino mass $m_{\nu_\mathrm{e}}$\footnote{The effective electron neutrino mass is defined as $m_{\nu_\mathrm{e}}^2=\Sigma |U_\mathrm{ei}|^2m_\mathrm{i}^2$, where $U_{\mathrm{ei}}$ are elements of the Pontecorvo-Maki-Nakagawa-Sakata matrix and $m_{\mathrm{i}}$ the eigenvalues of the neutrino mass eigenstates} by analyzing the shape of beta and electron capture (EC) spectra in the endpoint region relies only on energy and momentum conservation. Therefore, no fundamental physics assumptions should prevent this quantity from being determined, provided that experiments can achieve the required accuracy. 
This approach is followed by several collaborations that focus mainly on the beta decay of $^{3}$H and EC of $^{163}$Ho \cite{FORMAGGIO20211, Aker_2022, Esfahani_2017, Betti_2019, QTNM, ECHo_2014, Holmes}. The lowest upper limit on the effective electron anti-neutrino mass $m_{\overline{\nu_\mathrm{e}}}<0.45$ eV/c${^2}$ (90\% C.L.) has been achieved by the KATRIN collaboration with $^{3}$H \cite{KATRIN2024}, while the best limit on the effective electron neutrino mass $m_{\nu_\mathrm{e}}<27$ eV/c${^2}$ (90\% C.L.) has been achieved by the HOLMES collaboration \cite{Holmes_25}. 

The ECHo Collaboration \cite{ECHo_2014,ECHo2017} aims at determining the effective electron neutrino mass by studying the EC decay of $^{163}$Ho to $^{163}$Dy. The maximum energy available for $^{163}$Ho decay is $Q=2863.2(6)$ eV \cite{PTMS2024}. Thus electron capture can occur only starting from the third innermost electron shell. This, in turn, implies an extremely small fraction of photonic de-excitations, favoring de-excitation via Auger processes. To improve the sensitivity of the measurement and avoid systematic uncertainties, which occur in experiments with separated radioactive source and detector, the idea of enclosing the $^{163}$Ho source in high-energy resolution detectors was proposed in \cite{DERUJULA1982429} and is adopted in modern experiments \cite{ECHo_2014, Holmes}. The expected spectral shape in this type of measurements is, in first approximation, characterized by several resonances, each corresponding to an excited state in which the $^{163}$Dy atom may be left \cite{PhysRevC.97.054620}. The main resonances are MI at $2040$ eV corresponding to the capture of $3s$ electrons, MII at $1836$ eV for $3p_{1/2}$ electrons and NI at $411$ eV for $4s$ electrons.  
The key technology in ECHo is based on metallic magnetic calorimeters (MMCs) with enclosed $^{163}$Ho. MMCs are operated at temperatures below 20 mK and achieve an energy resolution in the eV range \cite{Fleischmann_2005,10.1063/1.3292407, KempfJLTP, Krantz25}. For MMCs optimized for the ECHo experiment, the fraction of events with partial energy detection in a region extending up to 50 eV below the EC endpoint energy $Q$ is lower than $2\times10^{-8}$, ensuring 100$\%$ detection efficiency in the endpoint region \cite{GASTALDO2013150, MANTEGAZZINI_2022166406}.  

\begin{figure}
\minipage{0.48\textwidth}
\begin{center}
\includegraphics[width=0.75\textwidth]{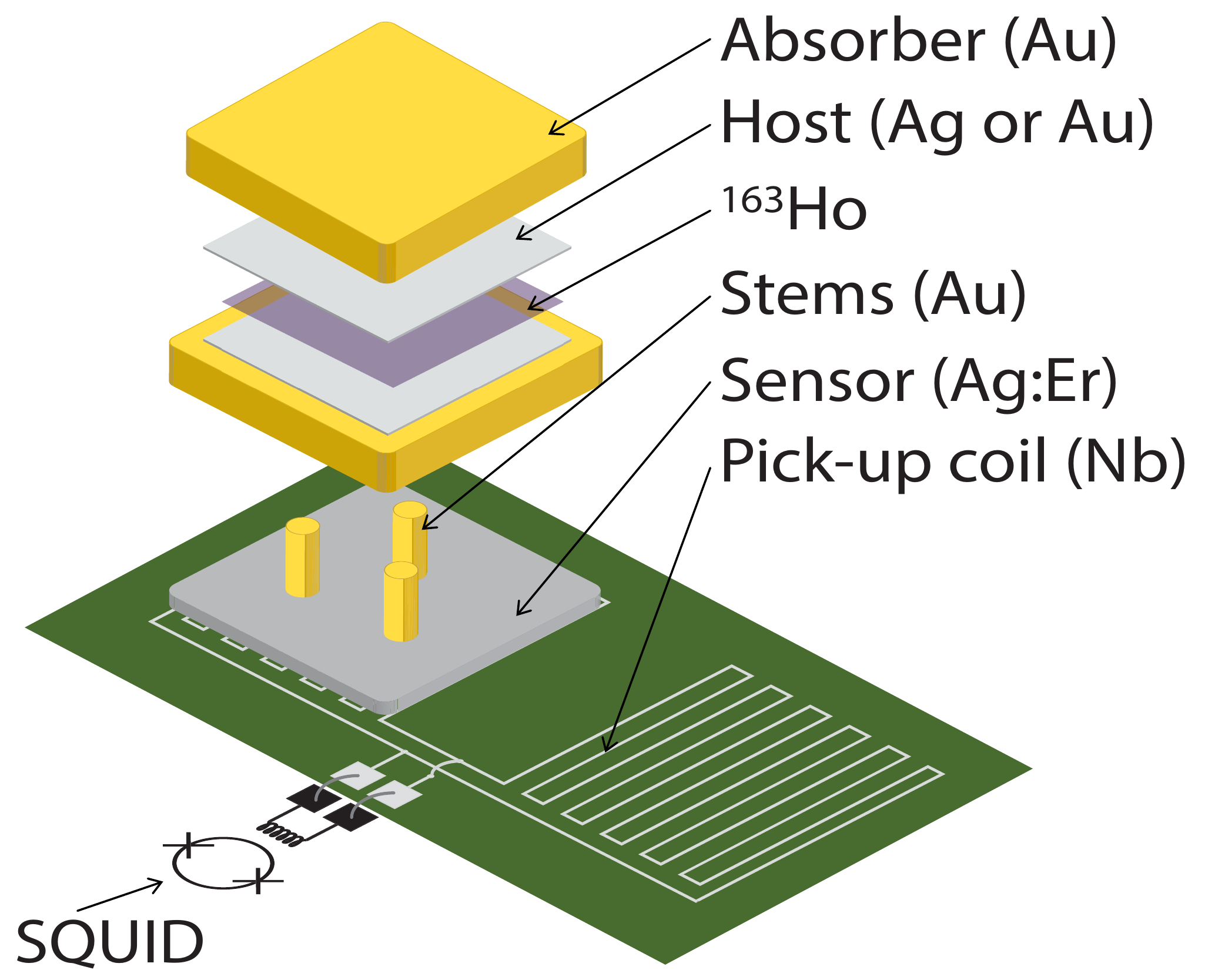}
\end{center}
\endminipage\hfill
    \caption{Schematic drawing of an MMC pixel for the ECHo experiment. For visual clarity, sensor and absorber are shown for only one of the two pixels, allowing the pick-up coil underneath to be seen.}
    \label{fig:1}
\end{figure}

The ECHo experiment has so far gone through two R\&D phases. In the first one, the possibility of obtaining a high-resolution $^{163}$Ho spectrum using MMCs \cite{Velte_EPJC} was demonstrated. Although only 4 MMC pixels with approximately 0.2 Bq $^{163}$Ho activity each were used, the ECHo collaboration set a new limit for the effective electron-neutrino mass $m_{\nu_\mathrm{e}} < 150$ eV/c${^2}$ (95\% C.L.) \cite{Velte_EPJC} improving the previous best upper limit of 225\,eV/c${^2}$ (95\% C.L.) \cite{PhysRevA.35.679}. The second phase, known as ECHo-1k, featured an optimized detector design \cite{MANTEGAZZINI_2022166406} and a new multi-channel readout scheme \cite{Mantegazzini_2021}. The $^{163}$Ho source was produced by irradiation of $^{162}$Er at the Institut Laue-Langevin in Grenoble and an optimized radiochemical separation ensured the required high purity \cite{Dorrer_Acta}. The ion-implantation procedure was performed at the upgraded RISIKO facility in Mainz \cite{Kieck_10.1063/1.5081094, KIECK2019162602}. 

This letter presents the analysis of the $^{163}$Ho EC spectrum measured in the ECHo-1k experiment with a total exposure of 4880 pixel $\times$ days and an average activity per pixel of 0.7 Bq \cite{MANTEGAZZINI_2022166406}.  
The success of the ECHo-1k experiment lays the foundation for the development of future large-scale experiments using $^{163}$Ho, with the aim of achieving sensitivity to the effective electron neutrino mass on par with experiments based on $^{3}$H and beyond.

For the ECHo-1k phase, a 72-pixel MMC array (36 channels), called ECHo-1k chip, has been developed \cite{MANTEGAZZINI_2022166406}. The MMC design is based on the meander-shaped pick-up coil geometry. The first order gradiometric configuration allows for the simultaneous readout of two pixels \cite{10.1063/1.3292407}. The single-pixel design for the ECHo-1k array has been optimized to achieve an energy resolution of $\Delta E_{\mathrm{FWHM}} < 10$ eV under realistic noise conditions. 
In figure \ref{fig:1} a schematic drawing of an MMC pixel of the ECHo-1k chip is shown. A meander-shaped niobium coil, micro-fabricated on a silicon substrate, has the dual function of carrying a persistent current, being niobium superconducting at the operating temperature of MMCs, and acts as a pick-up coil. A paramagnetic Ag:Er sensor is placed on top of the niobium structure and is thermally and mechanically connected to an absorber via three gold stems (for better visibility, sensor and absorber have been drawn only on one of the two sides of the coil). The persistent current running in the niobium coil generates a static magnetic field leading to a temperature-dependent magnetization of the paramagnetic sensor. The absorber exhibits a sandwich structure consisting of two gold layers. The volume of the first gold layer is $180\,\upmu \mathrm{m}  \times 180\,\upmu \mathrm{m} \times 5\,\upmu \mathrm{m}$. Into the top of this first gold layer, $^{163}$Ho ions are implanted within a central area of $150 \times 150 \,\upmu$m$^2$ which is covered with a $100$\,nm thin metallic layer (silver or gold). The implantation energy is 30 keV, corresponding to an average implantation depth of 5\,nm \cite{Gamer_2017}. A second gold layer with area $165 \times 165 \, \upmu$m$^2$ and thickness of $5\, \upmu$m is deposited on top of the first one. This second gold layer fully covers the implanted area and ensures complete containment of the radiation emitted in the EC of $^{163}$Ho. Ion implantation has been selected to enclose the $^{163}$Ho in particle absorbers since this process does not degrade the performance of detectors and allows the source atoms to be contained in a defined volume \cite{GASTALDO2013150}.
Two channels of the array, each equipped with just one sensor, are dedicated to measuring the temperature variations of the chip. The two pixels of these channels do not contain $^{163}$Ho. In addition, for seven channels, only one of the two pixels contains $^{163}$Ho to allow for in-situ background determination.
For the remaining 27 channels both pixels contain $^{163}$Ho. In the ECHo-1k experiment, two ECHo-1k chips have been used. The detectors in these chips are identical except for the material of the thin implantation layer, i.e. gold in one case (ECHo-1k-Au) and silver in the other case (ECHo-1k-Ag). The characterization of these two arrays has been discussed in \cite{MANTEGAZZINI_2022166406}. 
Each MMC channel is individually read-out using the two-stage dc-SQUID scheme \cite{Drung04, Drung_2006}. 
The pre-amplified voltage signals from MMC channels are digitized using 16-bit 125 MS/s ADC boards with synchronized clocks. The data acquisition software allows for online fitting of each trace passing the constant fraction trigger algorithm, providing guidance for the offline analysis. 
The temperature variation of the calorimeter following the energy release in the particle absorber has a pulse shape with time constants defined by the thermal properties of the detector. The amplitude is, in first approximation, directly proportional to the deposited energy.  
The acquired voltage signal conserves the pulse shape and the fairly linear dependence of the amplitude on the deposited energy. For the ECHo-1k detectors the non-linearity has been evaluated to be below 0.5\% at 2 keV and it is corrected with a quadratic energy calibration \cite{10.1063/1.3292407,MANTEGAZZINI_2022166406}.

In \cite{Hammann_EPJC} the basic concept of the analysis of ECHo data has been discussed. A two-step data reduction has been developed to discriminate single $^{163}$Ho events from triggered noise, pile-up events and events possibly related to cosmic muons or natural radioactivity. No meaningful energy dependence has been observed which implies no expected spectral distortions. Details on data reduction are given in Appendix \ref{A1}.
\begin{figure}
\minipage{0.48\textwidth}
\begin{center}
\includegraphics[width=0.95\textwidth]{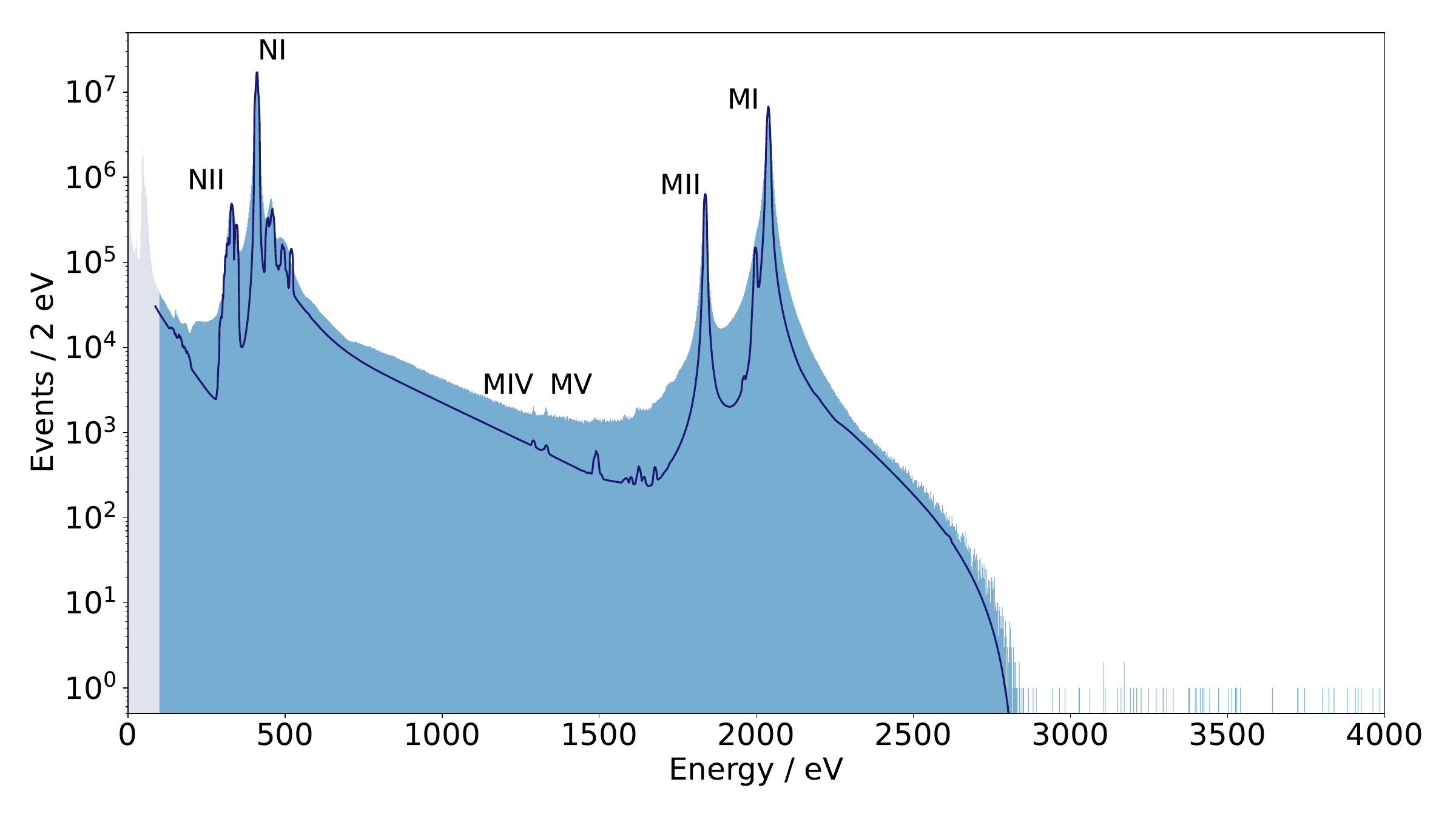}
\end{center}
\endminipage\hfill
    \caption{Calorimetrically measured $^{163}$Ho EC spectrum in the ECHo-1k experiment consisting of about 200 million events. The solid line is the output of a theoretical model \cite{Braß_2020} convoluted with a Gaussian having $\sigma = 2.8$ eV.}
    \label{Spectrum}
\end{figure}

Figure \ref{Spectrum} shows the measured $^{163}$Ho spectrum along with the convolution between the theoretical model derived in \cite{Braß_2020} and the Gaussian detector response (solid line). The model, based on an ab-initio approach, represents the most precise description of the electron capture process in $^{163}$Ho, but still fails in describing the tails and therefore cannot be used for fitting. The spectrum features about 200 million events between 100 eV and 5000 eV. The lower energy threshold is defined after correcting for trigger-level inhomogeneity among the channels. The effective energy resolution determined by the analysis of the MII-line, expected to be described by only one resonance, is $\Delta E_{\mathrm{FWHM}}\, =\, 6.59(16)$ eV. This value is then used as input parameter in the fit of the endpoint region. Other important parameters for the fit of the endpoint region of the $^{163}$Ho EC spectrum are the position and intrinsic width of the MI-line, modeled in first approximation as a simple Lorentzian, $E_{\mathrm{MI}}\, =\, 2039.7(5)\,$ eV and $\Gamma_{\mathrm{MI}}\, = \,14.0(2)$ eV. 
The unresolved pile-up fraction $f_{\mathrm{pu}}$, which defines the intensity of the pile-up background in the region of interest, is determined as the product of the activity per pixel and the rise time of the signal to be $f_{\mathrm{pu}}\, = \,0.8(5)\times 10^{-6}$. 
The fit of the endpoint region is then performed using a simplified analytical function for the measured decay rate:
\begin{equation}
\label{eq1}
\frac{dN}{dE} = C \times[ A(E) \times F_{\mathrm{PS}}(E,Q)] \otimes g(E,\sigma) + b(E)
\end{equation}
where $C$ is a constant. $A(E)$ describes the atomic physics contribution to the differential electron capture decay rate. In \cite{Braß_2020} a theoretical prediction of the differential electron capture nuclear decay rate is given. In the end point region, $A(E)$ is mainly determined by the 3$s$ electron capture followed by an Auger decay slightly modified by the sum over all possible decay channels and their interferences \cite{PhysRevC.97.054620}. The Auger matrix elements are energy dependent leading to a roughly Lorentzian line shape at the resonance and an exponential decay in the high energy tail \cite{Braß_2020}.
For the present analysis an analytical approximation needed to be defined. 
$F_{\mathrm{PS}}(E,Q)$ represents the part of the phase space function, which depends on the $Q$-value and on $m_{\nu_\mathrm{e}}$: $F_{\mathrm{PS}}(E,Q)=(Q-E)\times \sqrt{(Q-E)^2-m_{\nu_\mathrm{e}}^2}$. The term in squared brackets in Eq. \ref{eq1} is then convoluted with the Gaussian detector response $g(E,\sigma)$ with standard deviation $\sigma$ determined by fitting the MII-line. $b(E)$ is the background, which is composed of two contributions $b(E)= b_{\mathrm{const}}+ b_{\mathrm{pu}}$. $b_{\mathrm{const}}$ is an energy independent contribution due to natural radioactivity, which is in agreement with the expectation in \cite{Background_muon, Background_natural}. 
$b_{\mathrm{pu}}$ represents the unresolved pile-up contribution given by the normalized auto-convolution of the experimental spectrum weighted by the unresolved pile-up fraction $f_{\mathrm{pu}}$: $b_{\mathrm{pu}}=f_{\mathrm{pu}}(dN/dE\otimes dN/dE)$. The parameters $b_{\mathrm{const}}$ and $f_{\mathrm{pu}}$ have been determined by fitting the data in the energy region [2900, 5000] eV where only 80 events are present.  
\begin{figure}[t!]
\includegraphics[width=0.45\textwidth]{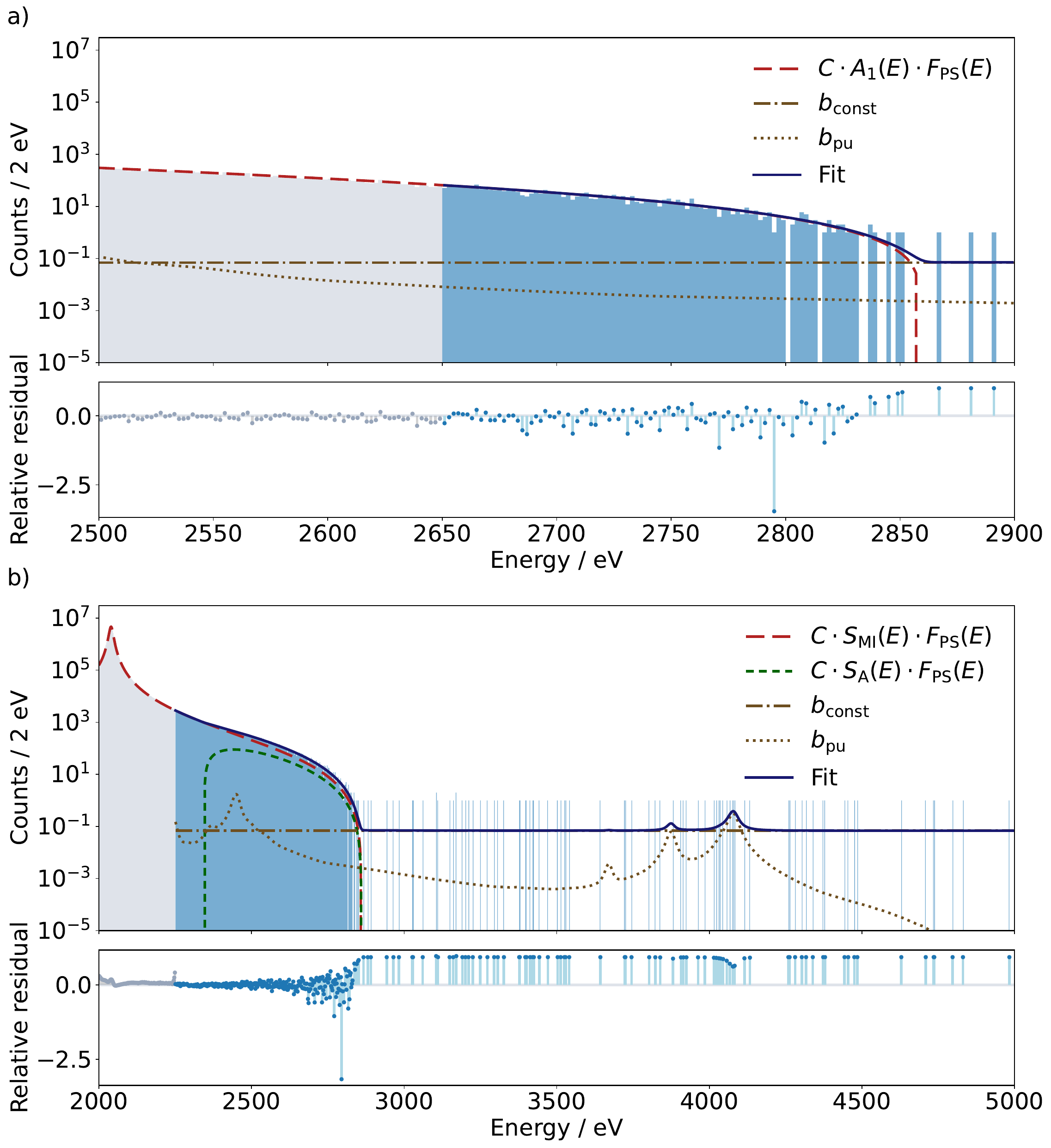}
\caption{Histogram of the endpoint region with superimposed fit function (blue solid line) with $A_{\mathrm{1}}$ (a) and $A_{\mathrm{2}}$ (b). For both plots, the fitting region is indicated with the blue histogram. The two brown lines, dotted and dash-dotted, represent the two background contributions $b_{\mathrm{pu}}$ and $b_{\mathrm{const}}$ respectively. The dashed lines represent the atomic physics functions multiplied with $F_{\mathrm{PS}}$.}
\label{fit}
\end{figure}
Two analytical functions are used for approximating $A(E)$. 
As discussed in \cite{Braß_2020} the function describing the atomic physics contribution is varying very slowly in the endpoint region. Thus, $A(E)$ can be approximated to a high level of accuracy by a simple exponential function in a small energy window below the end-point. The first analytical function has therefore the form: $A_{\mathrm{1}}(E) = A\times e^{-E/ \lambda}$. 
The parameters $A$ and $\lambda$ are determined in a grid fit of the spectrum in the range 2450 eV – 2750 eV using Eq. (\ref{eq1}) for a massless neutrino, Gaussian detector response, constant background and $Q= 2863.2$ eV \cite{PTMS2024}. 
In the second approach, the fitting range is increased towards smaller energies to reduce the statistical error on the fit parameters. 
The high energy tail of the MI-resonance, which dominates the spectral intensity in the endpoint region, is very complex. A simple Lorentzian, $S_{\mathrm{MI}}$, roughly describes the region around the maximum, whereby the tail shows, both experimentally and theoretically clear deviations from a Lorentzian line-shape. 
The approach used in \cite{Holmes_25} is followed and the deviation from a simple Lorentzian shape is described by adding a second function, $S_{\mathrm{A}}$, which is responsible for increasing the intensity of the spectrum towards the endpoint region:
\begin{equation}
\label{eq3}
\begin{split}
&A_{\mathrm{2}}(E) = k_{\mathrm{MI}}\times \frac{\frac{\Gamma_{\mathrm{MI}}}{2\pi}}{(E-E_{\mathrm{MI}})^2 + \frac{\Gamma_{\mathrm{MI}}^2}{4}} + \\
& k_{\mathrm{A}}\times \Theta (E-E_{\mathrm{A}})\times (e^{-(E-E_{\mathrm{A}})/ E_1}-e^{-(E-E_{\mathrm{A}})/ E_2}).    
\end{split}
\end{equation}
The constants $k_{\mathrm{MI}}$ and $k_{\mathrm{A}}$ indicate the intensity of the two contributions. $E_{\mathrm{A}}$ is the characteristic energy at which the additional contribution becomes important, while $E_1$ and $E_2$ are the energy scales on which the two exponential functions vary. 
The parameters describing $A_2(E)$ are first searched with a grid fit over the energy range $2250$ eV - $2750$ eV assuming massless neutrinos, fixed position and intrinsic width of the Lorentzian, fixed standard deviation for the Gaussian detector response, constant background and $Q= 2863.2$ eV \cite{PTMS2024}. 
The parameters describing the two $A(E)$ functions as derived with the grid fits are used to perform a Bayesian inference on the histogram of the endpoint region of the $^{163}$Ho EC spectrum. The aim is to simultaneously estimate posterior distributions of the spectral shape parameters, $Q$-value and neutrino mass to best explain the data under the assumption of Poisson statistics.
Using the Stan statistical software platform \cite{stan}, this estimation was accomplished through the Hamiltonian Monte Carlo (HMC) algorithm, specifically using the implementation of the No-U-Turn Sampler (NUTS). 
The parameters describing the $^{163}$Ho EC spectrum and the constant background are included with a normal prior, while for the effective neutrino mass a flat prior for $0$ eV $<$ $m_{\nu_\mathrm{e}}$ $<$ $100$ eV is used. 
Non-informative priors are used in case of $A_{\mathrm{2}}$ while for $A_{\mathrm{1}}$ all priors, besides the effective neutrino mass, need to be constrained to the grid uncertainties due to a large correlation among them. This is due to the presently relatively small number of counts in the fit region.

\begin{figure}[t!]
\includegraphics[width=0.45\textwidth]{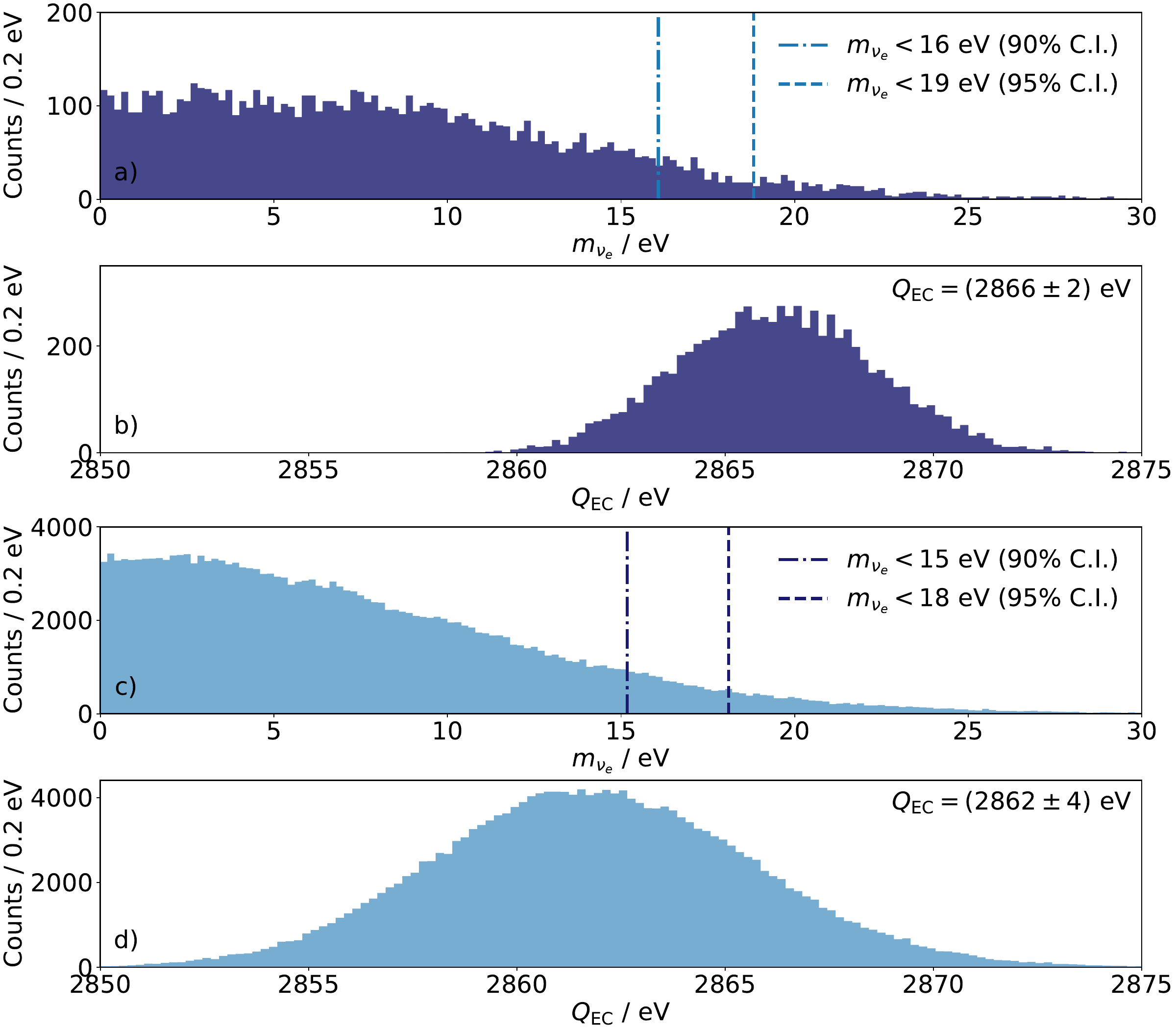}
\caption{Posterior distribution for $Q$ and $m_{\nu_\mathrm{e}}$ calculated for $A_{\mathrm{1}}$ (a and b respectively) and for $A_{\mathrm{2}}$ (c and d respectively).}
\label{stan}
\end{figure}

Figure \ref{fit} a) shows the histogram of the endpoint region of the $^{163}$Ho EC spectrum with the fit function for $A_{\mathrm{1}}$ as blue solid line (fitting range 2650 eV - 2900 eV). The red-dashed line represents $C\, \times \, A_{\mathrm{1}} \,\times\, F_{\mathrm{PS}}$.
Figure \ref{fit} b) shows the histogram in a larger range with the fit function for $A_{\mathrm{2}}$ as blue solid line (fitting range: 2250 eV - 5000 eV). The red-dashed line represents the product of $S_{\mathrm{MI}}$ and $F_{\mathrm{PS}}$ while the green short-dashed line represents the product of $S_{\mathrm{A}}$ and $F_{\mathrm{PS}}$. The single background components, are shown in both plots ($b_{\mathrm{const}}$ brown dash-dotted line and $b_{\mathrm{pu}}$ brown dotted line). The constant background derived in both fits is the same: $0.073(8)$ /2eV, which, given an effective exposure of 4000 pixel $\times$ day, leads to a background rate of $B=9.1(1.3)\times 10^{-6}$ /eV/pixel/day. 
The posterior distributions for $Q$ and $m_{\nu_\mathrm{e}}$ are shown in Figure \ref{stan} for $A_{\mathrm{1}}$ (plots a) and b)) and $A_{\mathrm{2}}$ (plots c) and d)). The $Q$-value derived from the posterior distribution for $A_{\mathrm{1}}$ is $Q = 2866.3(2.0)$ eV (error related to the narrow prior) and the neutrino mass limit is $m_{\nu_\mathrm{e}} < 19\, $ eV at 95\% credible interval (C.I.) ($m_{\nu_\mathrm{e}} < 16\, $ eV at 90\% C.I.). The corresponding values for $A_{\mathrm{2}}$ are $Q = 2862(4)$ eV and $m_{\nu_\mathrm{e}} < 18\, $ eV at 95\% C.I. ($m_{\nu_\mathrm{e}} < 15\, $ eV at 90\% C.I.). The results obtained using the two different models for the spectral shape are compatible, nevertheless the use of narrow priors in case of $A_{\mathrm{1}}$ makes the derivation of $Q$ and $m_{\nu_\mathrm{e}}$ dependent on the parametrization. This, in turn, indicates that the use of a simple exponential function at the present level of statistics, is not appropriate. This approach might become important once the total number of counts is at least one order of magnitude higher. 

We have presented the analysis of the $^{163}$Ho EC spectrum acquired within the ECHo-1k experiment consisting of about 200 million events. 
Two analytical functions have been tested to describe the endpoint region of the $^{163}$Ho EC spectrum in the absence of a complete theoretical model, $A_{\mathrm{1}}$ and $A_{\mathrm{2}}$. At the present level of statistics the simplified function $A_{\mathrm{1}}$ does not properly represent the data.
Using $A_{\mathrm{2}}$ the $Q$-value was derived to be $Q = 2862(4)$ eV, which is in excellent agreement with the one determined via Penning-trap mass spectrometry \cite{PTMS2024}. The upper limit of the effective electron neutrino mass could be reduced to $m_{\nu_\mathrm{e}} < 15\, $ eV (90\% C.I.). 
This limit is in agreement with the expectation given the available statistics and the achieved very low background rate of $B=9.1(1.3)\times 10^{-6}$ /eV/pixel/day. This new upper limit represents an improvement of almost one order of magnitude with respect to the value in \cite{Velte_EPJC} and of almost a factor of 2 with respect to the recent limit in \cite{Holmes_25}. Along with the demonstrated multiplexing readout of MMC arrays \cite{Richter} and the possibility to increase the activity per pixel \cite{Griedel}, these results set the basis for the next stage of the ECHo experiment. 
This new upper limit on the effective electron neutrino mass together with the results in \cite{Holmes_25} strengthen the role of the $^{163}$Ho-based experiments ECHo and HOLMES for a model-independent determination of the neutrino mass scale. In turn, this paves the way for comparing the value of the effective electron neutrino and anti-neutrino mass from future large scale experiments for CPT violation test. 

We acknowledge funding and support by the Deutsche Forschungsgemeinschaft (DFG, German Research 
Foundation) Research Unit FOR2202 Neutrino Mass Determination by Electron Capture in $^{163}$Ho, ECHo (funding under grant numbers HA 6108/2 (M.B., M.W.H.), GA 2219/2 (A.B., D.B., C.E., L.G., F.M., C.V.), EN 299/7 (C.E., S.K., M.W.), EN299/8 (C.E.), BL 981/5 (K.B., S.E., R.X.S., C.S.) and DU 1334/1 (C.E.D., H.D., K.W.)). F.M. and A.B. acknowledge the support by the Research Training Group HighRR funded by the DFG. M.W.H., K.B., S.E., P.F. and M.D. acknowledge the support by the DFG through SFB 1225 ISOQUANT Project-ID 273811115.
M.N. acknowledges financial support by the Karlsruhe School for Elementary Particle and Astroparticle Physics: Science and Technology (KSETA)“. L.E.A.P., R.G., T.M., O.S. and M.W. acknowledge the financial support by Helmholtz. D.H. and A.R. acknowledge support by the European Microkelvin Platform (EMP). 
We acknowledge the support of the cleanroom team, the Electronics Workshop and the Mechanical Workshop at the Kirchhoff Institute at Heidelberg University and the nuclear chemistry and radiation protection team at the Department of Chemistry and the TRIGA staff at Mainz University (N. Bittner, C. Gorges, C. Mokry, D. Renisch, J. Riemer, J. Runke, L. Winkelmann).  The authors gratefully acknowledge the data storage service SDS@hd supported by the Ministry of Science, Research and the Arts Baden-Württemberg (MWK) and the DFG through grant INST 35/1503-1 FUGG.

\appendix
\section{Data Processing}
\label{A1}
The first step, called the Time-Info-Filter (TIF), contains algorithms selecting events on the basis of the trigger time. The filters implemented in this step are: the coincidence filter to eliminate events when more than one detector channel is triggered within the coincidence time, the holdoff filter to eliminate events which occur on the tail of a previous event within the same channel and the burst filter which eliminates events in one channel if the rate in a given time interval exceeds a threshold defined based on the $^{163}$Ho activity in that channel. 
The fraction of triggered traces identified as bad events by the TIF is 24.9$\%$ for the pixels in ECHo-1k-Ag and 5$\%$ for the pixels in ECHo-1k-Au, in both cases dominated by small-amplitude electromagnetic noise, as caused by cellular phones \cite{Hammann_EPJC}. The higher exclusion fraction for ECHo-1k-Ag pixels is due to suboptimal shielding of the readout cables. This large fraction of excluded events, in particular through the burst filter, leads to a reduced effective exposure of about 4000 pixel$\times$day. 

In the second data reduction step, events surviving the TIF are analyzed to identify and discard events whose shape differs from the expected pulse shape, obtained as average of a large number of $^{163}$Ho MI-signals. For the analysis of ECHo-1k data a template fit method has been selected and the $\chi ^2$ of the trace fit is used to filter out events such as pile-up and triggered noise not identified with the TIF. The $\chi ^2$ selection region is defined on the histogram of the $\chi ^2$ values as $\pm 3\sigma$ region around the maximum, assuming Gaussian distribution.  
In order to identify the leakage of possible spurious events in the region of interest of the $^{163}$Ho spectrum above 2.5 keV, the $\chi ^2$ histograms are analyzed for different low threshold energies. A leakage contribution well below the statistical uncertainty has been estimated for the endpoint region. 
After the $\chi ^2$ filter, 4.3$\%$ and 5.9$\%$ of the data surviving the TIF are discarded for ECHo-1k-Ag and ECHo-1k-Au, respectively. The major part of the discarded events, about 50$\%$, has an energy below 100 eV and represents either low amplitude triggered noise or change of slope of the baseline due to slight temperature fluctuations of the substrate, inducing a trigger. The fraction of discarded events at higher energy is in agreement with the expected number of pile-up events. 
To evaluate a possible energy dependence of the applied filters, the ratio between the number of events in the NI-line and the number of events in the MI-line was analyzed. Datasets having this ratio outside five times the standard deviation are discarded. The fraction of removed events with respect to the events surviving the $\chi ^2$ filter is 8$\%$ for the ECHo-1k-Ag data, corresponding to measurements in which the working point of the SQUID readout was not stable. Similarly, the ratio between the number of events in the MI-line and the number of events in the region 2.5 keV – 2.9 keV was monitored. No events among those surviving all previous cuts were discarded. 
 
After pulse selection, two steps are still necessary before obtaining the energy spectra for each pixel. The first one is 
a correction for amplitude variations due to temperature variations of the chip. This correction is obtained by correlating the amplitude of events in the MI line with the voltage output of the two temperature channels on the chip. 

The second and final step is the energy calibration using the resonances present in the $^{163}$Ho EC spectrum. 
Already with the very first MMCs with implanted $^{163}$Ho, the ECHo collaboration has pointed out that the shape of the $^{163}$Ho spectrum is more complex than foreseen \cite{PhysRevLett.119.122501}. This observation motivated several works proposing models describing the EC in $^{163}$Ho \cite{Faessler_2015, PhysRevC.91.035504,  PhysRevC.91.045505, PhysRevC.91.064302, PhysRevC.95.045502, ADRML2016, PhysRevC.97.054620, Braß_2020}. The theoretical calculations, which best describe the observed spectral features are based on an ab-initio approach \cite{PhysRevC.97.054620, Braß_2020}. 
For this reason, after the ECHo-1k data acquisition, a new calibration measurement was performed. In this experiment a $^{55}$Fe source was used for the calibration of the $^{163}$Ho spectrum acquired with pixels in the ECHo-1k-Ag array featuring an energy resolution below 5 eV FWHM. The determination of the line position was performed with a symmetric Gaussian fit for the MII-line and an asymmetric Gaussian fit for the NI and MI lines, to account for the multiplet nature of those lines.  
The positions of the $^{163}$Ho lines obtained with respect to the quadratic calibration function based on the $^{55}$Fe lines do not differ much from the value in \cite{PhysRevLett.119.122501}: MI $2040.67(11)$ eV, MII $1836.55(11)$ eV and NI $411.81(8)$ eV, but are more accurate.
The same fitting range to extract the position of the $^{163}$Ho lines in the calibration analysis is then applied to each single spectrum and a quadratic function is used for calibrating the data. The parameters of the quadratic function for each pixel do not appreciably vary over the measurement time. As control procedure for the robustness of the calibration, the position of the calibrated lines is compared to the expected positions. The average deviation for the position of the calibrated lines is around $0.01\%$.

The determination of the energy resolution in the calibrated spectra for each pixel and for each measurement is then performed. This is estimated from the width of the NI-line. This line was chosen over the others despite its more complex line shape because it has the highest number of events with respect to the other lines.
The distribution of the energy resolution as FWHM has a Gaussian shape giving an average value of $\Delta E_{\mathrm{FWHM}} =7.2(5)$ eV for pixels in the ECHo-1k-Ag and $\Delta E_{\mathrm{FWHM}} =5.2(3)$ eV for pixels in the ECHo-1k-Au. The effect of having single spectra with different energy resolution composing the high statistics $^{163}$Ho EC spectrum on the derivation of the parameters describing the endpoint region has been studied through Monte Carlo simulations assuming massless neutrinos. 
The fit of the end-point region, applying algorithms described in the main text, provided $Q$-values in agreement with the one fixed for the simulation within the statistical uncertainty. This ensures that the systematic error induced by ignoring the slight differences in detector response of the used pixels is still smaller than the statistical error.


\nocite{*}

\bibliography{apssamp}

\end{document}